\documentclass[aps,pra,showpacs%
        ,amsmath%
        ,superscriptaddress%
        ,reprint%
        ,floatfix
        ,nofootinbib
        ,preprintnumbers
        ]{revtex4-1}

\usepackage[pdftex]{graphicx} 

\usepackage{amsmath,amssymb}

\newcommand{\paulivec}{\bvec{\sigma}}
\newcommand{\bbset}[1]{\mathbb{#1}}
\newcommand{\Real}{\bbset{R}}
\newcommand{\set}[1]{\left\{ #1 \right\}}

\newcommand{\bvec}[1]{\boldsymbol{#1}} 

\newcommand{\MP}{\mathcal{P}} 
\newcommand{\MM}{\mathcal{M}} 

\newcommand{\RPP}{\Real{}P^2} 

\begin{document}

\title{Path topology dependence of adiabatic time evolution}

\author{Atushi Tanaka}
\homepage[]{\tt http://researchmap.jp/tanaka-atushi/}
\affiliation{Department of Physics, Tokyo Metropolitan University, Hachioji, Tokyo 192-0397, Japan}

\author{Taksu Cheon}
\homepage[]{\tt http://researchmap.jp/T_Zen}
\affiliation{Laboratory of Physics, Kochi University of Technology, Tosa Yamada, Kochi 782-8502, Japan}

\begin{abstract}
An adiabatic time evolution of a closed quantum system connects eigenspaces of initial and final Hermitian Hamiltonians for slowly driven systems, or, unitary Floquet operators for slowly modulated driven systems. We show that the connection of eigenspaces depends on a topological property of the adiabatic paths for given initial and final points. An example in slowly modulated periodically driven systems is shown. These analysis are based on the topological analysis of the exotic quantum holonomy in adiabatic closed paths.
\end{abstract}




\maketitle

\section{Introduction}
The adiabatic theorem for isolated quantum systems is a basic 
principle
of the quantum dynamics: Once the system is prepared to be in a
stationary state, the system remains to be stationary as long as
the parameters
of the system
are
varied slow enough. There are many proof for
slowly driven 
systems~\cite{Born-ZP-51-165,Kato-JPSJ-5-435,Avron-CMP-110-33}, 
which are described by Hermitian Hamiltonian, as well as slowly
modulated driven 
systems~\cite{Young-JMP-11-3298,Dranov-JMP-39-1340,Tanaka-JPSJ-80-125002}, 
which are described by unitary Floquet operators.
The adiabatic theorem has diverse applications, e.g. in 
molecular science and solid state physics~\cite{Born-AnnPhysik-389-457,BornHuang-DTCL-1954},
quantum holonomy~\cite{Shapere-GPP-1989,Bohm-GPQS-2003,Chruscinski-GPCQM-2004}
and adiabatic quantum computation~\cite{Farhi-quant-ph-0001106,Farhi-Science-292-472,Tanaka-PRA-81-022320}.

In this article, we examine how the final stationary state of the
adiabatic time evolution depends on the path in the adiabatic parameter
space.
In particular, 
we here focus on the eigenspaces corresponding to the initial and final
stationary states,
and we will ignore the phase information in the following.
First, we will show that the final stationary state generally depends 
on the adiabatic path, although the initial adiabatic parameter and 
initial stationary state are kept fixed. 
It turns out that 
the topology of the adiabatic path plays the key role there.
Second, we will show that the discrepancy between two final stationary 
states corresponding to two different adiabatic path is characterized
by a permutation matrix, which is governed by a homotopy equivalence.
Our idea is an application of the topological formulation for
the exotic quantum holonomy~\cite{Cheon-PLA-248-285,Tanaka-PRL-98-160407}, 
which concerns the nontrivial change
in eigenspaces induced by adiabatic cycles~\cite{Tanaka-PLA-379-1693}.

The present argument heavily relies on topology, in particular the concept of homotopy and its application to the covering map. At the same time, our argument is formal in the sense of mathematics. Since our argument relies only on an elementally account of homotopy and covering map, we refer textbooks of topology for more mathematical description~\cite{Lee-ITT-2011,Hatcher-AT-2001,Nakahara-GTP-1990}.
The covering map is also discussed in a study of phase holonomy of non-Hermitian quantum systems~\cite{MehriDehnavi-JMP-49-082105}.

The plan of this article is the following.
In Section~2, we introduce the lifting of adiabatic paths for our problem. 
This is considered to be an extension of the lifting for 
the phase holonomy~\cite{Simon-PRL-51-2167,Aharonov-PRL-58-1593}.
In Section~3, we present the main results.
An example is shown in Section~4.
We conclude this article in Section~5.

\section{Lifting adiabatic paths}
The lifting of adiabatic paths is the central concept for the theory of conventional quantum holonomy~\cite{Shapere-GPP-1989,Bohm-GPQS-2003,Chruscinski-GPCQM-2004}. We take over this concept to examine the path-dependence of eigenspaces. In this sense, our approach is a straightforward extension of the works by Simon~\cite{Simon-PRL-51-2167} and Aharonov and Anandan~\cite{Aharonov-PRL-58-1593}.

We here focus on the simplest case where the system is described by
a $N$-level Hermitian Hamiltonian $H(\lambda)$ with an adiabatic 
parameter $\lambda$, whose space is denoted as $\MM$.
The energy spectrum of $H(\lambda)$ is 
assumed to be discrete and non-degenerate 
for an arbitrary $\lambda$ in $\MM$.
Let $P_n(\lambda)$ be the eigenprojector corresponding to 
an eigenenergy $E_n(\lambda)$ ($n=0,1,\ldots,N-1$).
Our assumption on the eigenenergies ensures that 
$E_n(\lambda)$ and $P_n(\lambda)$ are smooth in $\MM$.
Also, $P_n(\lambda)$ is rank-one.
We remark that it is straightforward to extend the following analysis 
to unitary Floquet systems, if the spectrum of the Floquet operator is 
discrete and non-degenerate.

We examine all eigenprojectors at a time, 
which facilitates to compare the changes in eigenspaces induced
by two adiabatic paths.
From the eigenprojectors $P_n(\lambda)$ for a given point $\lambda$ in $\MM$,
we introduce an ordered set of the eigenprojectors
\begin{align}
  \label{eq:p_def}
  p(\lambda) 
  \equiv \left({P}_0(\lambda), {P}_1(\lambda), 
    \ldots,{P}_{N-1}(\lambda)\right)
  .
\end{align}
We may introduce another ordered sets, since the order is arbitrary.
Let $\sigma$ denote a permutation of quantum numbers $0,1,\ldots,N-1$.
In other words, $\sigma$ is an element of $N$-th symmetric group ${\cal S}_N$.
Let $p_{\sigma}(\lambda) = \sigma\left(p(\lambda)\right)$, i.e.,
\begin{align}
  \label{eq:psigma_def}
  p_{\sigma}(\lambda) 
  \equiv \left({P}_{\sigma(0)}(\lambda), {P}_{\sigma(1)}(\lambda), 
    \ldots,{P}_{\sigma(N-1)}(\lambda)\right)
  ,
\end{align}
where $\sigma(n)$ is the permutated quantum number 
for a given quantum number $n$.

We introduce a fiber at $\lambda$ in $\MM$:
\begin{equation}
  \label{eq:defFlambda}
  F_{\lambda}\equiv
  \bigcup_{\sigma\in{\mathcal{S}_N}}\set{p_{\sigma}(\lambda)}
  .
\end{equation}
For an arbitrary pair of elements, say $p_{\sigma'}(\lambda)$ and
$p_{\sigma''}(\lambda)$, of $F_{\lambda}$,
there is a unique permutation $\sigma\in\mathcal{S}_N$ that satisfies
$p_{\sigma''}(\lambda) = \sigma\left(p_{\sigma'}(\lambda)\right)$,
i.e., $\sigma''(n) = \sigma(\sigma'(n))$ for an arbitrary $n$.
In this sense, we call $\mathcal{S}_N$ a structural group
of the fiber $F_{\lambda}$.

A total space $\MP$ consists of the fibers $F_{\lambda}$ in $\MM$:
\begin{equation}
  \label{eq:defMP}
  \MP \equiv \bigcup_{\lambda\in\MM} F_{\lambda}
  , 
\end{equation}
which naturally accompanies a projection 
\begin{equation}
  \label{eq:pi}
  \pi: \MP \to \MM
\end{equation}
by construction.
Hence we obtain a fiber bundle consists of 
the total space $\MP$, the projection $\pi$, 
the base manifold $\MM$ and the fiber $F_{\lambda}$.

Utilizing the fiber bundle introduced above, we introduce a lifting of a path $C$ in $\MM$ to $\MP$ in order to examine the adiabatic time evolution of $p$ along $C$. Let $\lambda_{\rm i}$ and $\lambda_{\rm f}$ denote the initial and final points of $C$, respectively.

The adiabatic time evolution of $p\in\MP$ can be determined by the time evolution of each eigenprojector $P_n$.
The adiabatic theorem ensures that the final state of the adiabatic time evolution along $C$ is unique
for a given initial stationary state $P_n(\lambda_{\rm i})$.
Accordingly the initial $p_{\rm i}\in F_{\lambda_{\rm i}}$ and the adiabatic path $C$ uniquely determines the final point $p_{\rm f} \in F_{\lambda_{\rm f}}$. Let $\phi_C$ from $F_{\lambda_{\rm i}}$ to $F_{\lambda_{\rm f}}$ denote the mapping from the initial to the final point, i.e., 
\begin{equation}
  \label{eq:def_phi_C}
  p_{\rm f} = \phi_C(p_{\rm i}) 
  .
\end{equation}
Namely, the mapping $\phi_C$ describes the change of $p$ induced by the adiabatic path $C$.
\begin{figure}[h]
  \begin{center}
  \includegraphics[width=5cm]{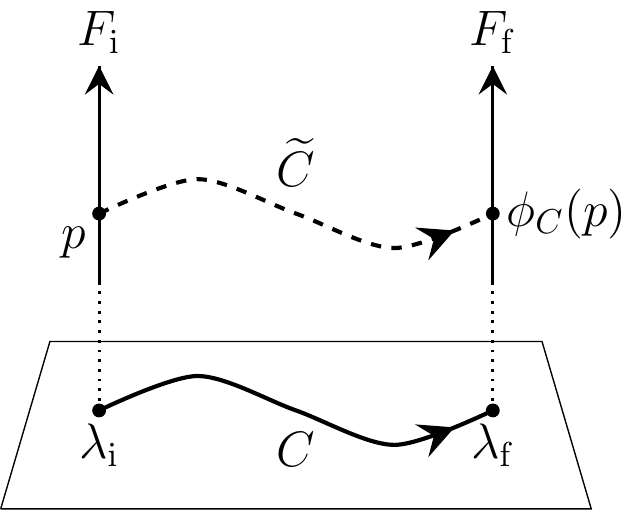}
  \end{center}
  \caption{%
    Lifting adiabatic path $C$ in $\MM$ to $\MP$ (Eq.~\eqref{eq:defMP}),
    which is made of fibers $F_{\lambda}$ (Eq.~\eqref{eq:defFlambda}).
    Let $\lambda_{\rm i}$ and $\lambda_{\rm f}$ denote the initial
    and final point of $C$, respectively.
    The lifted path $\tilde{C}$ starts from $p$, which is 
    in the initial fiber $F_{\rm i}$, and satisfies the adiabatic
    Schr\"odinger equation for the ordered set of eigenprojectors.
    We introduce the mapping $\phi_C$ from $F_{\rm i}$
    to the final fiber $F_{\rm f}$,
    so that $\phi_C(p)$ is the final point of $\tilde{C}$.
  }
\label{fig:lift}
\end{figure}

\begin{figure}[h]
  \centering
  \includegraphics[width=4.3cm]{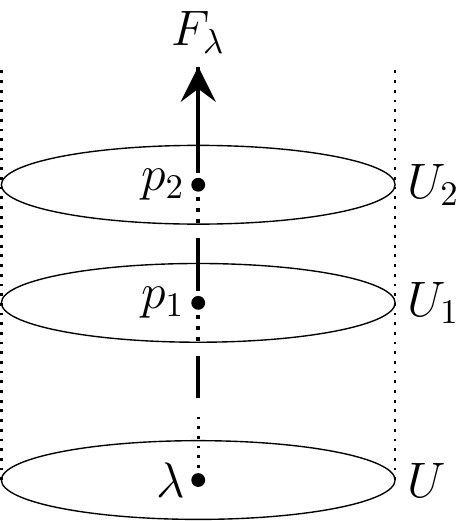}
  \caption{%
    A schematic picture of the covering map $\pi: \MP\to\MM$ 
    (Eq.~\eqref{eq:pi}).
    Let $\lambda$ be a point in an open set $U\subset\MM$.
    Points $p_j$ in the fiber $F_{\lambda}$ (Eq.~\eqref{eq:defFlambda})
    satisfies $\pi(p_j)=\lambda$.
    When $\pi$ is a covering map, 
    $\pi^{-1}(U)$ consists of disjoint union of open sets $U_j$,
    each of which is mapped homeomorphically onto $U$.
  }
  \label{fig:covering}
\end{figure}

The projection $\pi$ introduced in Eq.~\eqref{eq:pi} satisfies the axiom of covering projections~\cite{Lee-ITT-2011}.
Namely, for a given point $\lambda$ in $\MM$, there is an open
subset $U$ of $\MM$ that satisfies the following: 
$\pi^{-1}(U)$ is a disjoint union of connected open subset
of $\MP$. 
Each of the disjoint component $U_j$ is mapped homeomorphically onto $U$ 
(Fig.~\ref{fig:covering}).
The covering map structure determines
a various properties of $\phi_C$. In particular, it will be shown
below that the homotopic classification of the paths plays 
the central role here.

If $C$ is a closed path with a given
initial point $\lambda_{\rm i}$, the mapping $\phi_{C}$ on 
$F_{\lambda_{\rm i}}$ is called the monodromy action
(Theorem 11.22 in Ref.~\cite{Lee-ITT-2011}).
We also note that $\phi_{C}$ can be regarded as a permutation of 
eigenprojector at $\lambda=\lambda_{\rm i}$.
Since our argument in the next section much owes to the properties of the 
monodromy action (e.g. shown in Ref.~\cite{Lee-ITT-2011}),
we will quote the relevant result where appropriate.

\section{Comparison of adiabatic paths}
We lay out our main results in this section. 
We compare two adiabatic paths $C_1$ and $C_2$, which have the same 
initial and final points $\lambda_{\rm i}$ and $\lambda_{\rm f}$, 
in the adiabatic parameter space $\MM$.
For a given initial eigenprojector at $\lambda_{\rm i}$, 
we will elucidate how the eigenprojectors at the $\lambda_{\rm f}$ depends
on $C_1$ and $C_2$, 
by examining the adiabatic time evolutions of the ordered set of 
eigenprojectors (Eq.~\eqref{eq:p_def}).
In other words, we examine how $\phi_{C}$, which is a mapping
from $F_{\lambda_{\rm i}}$ to $F_{\lambda_{\rm f}}$,
depends on the topology 
of the path.

First of all, we examine the case that $C_1$ is homotopic to $C_2$, which is denoted as $C_1\sim C_2$. Namely, we suppose that we may smoothly deform $C_1$ to $C_2$, while keeping its initial and final points. 
We remark that this is the case where most conventional analyses of the periodic adiabatic time evolution have focused on.

If $C_1$ and $C_2$ are homotopic, $\phi_{C_1}$ and $\phi_{C_2}$
are identical, due to the homotopy lifting property 
(e.g., Theorem 11.13 in Ref.~\cite{Lee-ITT-2011}).
Hence, an arbitrary initial eigenprojector is adiabatically transported 
to the same final point through the paths $C_1$ and $C_2$.

Utilizing this result, we denote by $\phi_{[C]}$, instead of $\phi_C$, where 
$[C]$ denotes the equivalence class of a path $C$ under the homotopic 
classification.

Secondly, we proceed to the case where $C_1$ is not homotopic to $C_2$.
We compare these paths with 
a closed path $C\equiv C_1\cdot(C_2)^{-1}$,
where the inverse path of $C_2$ follows after $C_1$.
Hence the initial point of $C$ is $\lambda_{\rm i}$.
If a closed path $\gamma$, whose initial point is $\lambda_{\rm i}$,
in $\MM$ is homotopic to $C$, 
the following formula for $\phi_{[C]}$ holds:
\begin{equation}
  \label{eq:compositePhi}
  \left(\phi_{[C_2]}\right)^{-1}\circ\phi_{[C_1]}=\phi_{[\gamma]}
  ,
\end{equation}
where $\circ$ denotes the composition of the mappings $\phi_C$.
Eq.~\eqref{eq:compositePhi} is shown in the following way.
Because of $C\sim\gamma$, $\phi_{[C]}=\phi_{[\gamma]}$ holds.
On the other hand, 
$\phi_{[C]} 
= \left(\phi_{[C_2]}\right)^{-1}\circ\phi_{[C_1]}$ holds
from the definition of $\phi_{[C_1\cdot(C_2)^{-1}]}$ 
(Eq.~\eqref{eq:def_phi_C}).

Now our problem, i.e., the comparison of adiabatic time evolutions
along adiabatic paths in $\MM$, 
is casted into the analysis of the monodromy action $\phi_{[\gamma]}$
for an arbitrary closed path $\gamma$ in $\MM$.
We remind that 
$\phi_{[\gamma]}$ 
corresponds to a permutation of eigenprojectors induced by
the adiabatic time evolution along $\gamma$.
For example, if $\gamma$ is contractable to 
the point $\lambda_{\rm i}$, 
which is equivalent to the case $C_1\sim{}C_2$ examined above,
$\phi_{[\gamma]}$ corresponds to the identical permutation,
which implies $\phi_{[C_1]} = \phi_{[C_2]}$.

In order to completely solve our problem,
there are two tasks. One is to enumerate all equivalence class
$[\gamma]$ of closed paths in $\MM$.
Namely, we need to identify the first fundamental group $\pi_1(\MM)$
of the adiabatic parameter space $\MM$.
The other is to examine the monodromy action $\phi_{[\gamma]}$ of 
eigenspaces, for every $[\gamma]$ in $\pi_1(\MM)$.

There remains a question whether different equivalent
classes $[\gamma]$ and $[\gamma']$ induce different permutations of
eigenspaces. In other words, we need to clarify whether 
$\pi_1(\MM)$ completely characterizes 
the collection of $\phi_{[\gamma]}$.
There are two cases:
\begin{enumerate}
\item If $\MP$ is simply connected, i.e., $\pi_1(\MP)$ has only a
single element, $\phi_{[C_1]} = \phi_{[C_2]}$ holds
if and only if $C_1$ is homotopically equivalent to $C_2$. 
Hence $\pi_1(\MM)$ offers the complete
classification of the adiabatic cycles for our problem.
\item
In general, we need to modify the first case above, where 
the equivalence class of closed paths $\pi_1(\MM)$
is replaced with $H$ where $H\equiv\pi_1(\MM)/\pi_*(\pi_1(\MP))$.
Namely, $\phi_{[C_1]} = \phi_{[C_2]}$ holds
if and only if $C_1$ is equivalent to $C_2$ under the equivalence class
$H$ of closed paths in $\MM$.
\end{enumerate}
Here we assume that $\pi$ is a normal covering map, which is 
equivalent to the condition that $H$ is independent of $p_{\rm i}$
(Proposition 11.35 in Ref.~\cite{Lee-ITT-2011}). As far as we see, this assumption holds in 
our examples.

These result concern with the group $\Phi$ consists of 
all possible $\phi_{[\gamma]}$ for an arbitrary closed path $\gamma$.
In the theory of covering map, $\Phi$ is called a covering automorphism 
group, and the above result is just the one-to-one correspondence
between $\Phi$ and $H$ (Theorem 12.7 in Ref.~\cite{Lee-ITT-2011}).

We examine the latter, general case, where $\MP$ is multiply connected.
There is a closed path $\tilde{C}$ that is not contractable
to a point, in $\MP$. We assume that the initial point 
$p_{\rm i}$ of $\tilde{C}$ satisfies $\pi(p_{\rm i})=\lambda_{\rm i}$.
Let $C$ be the projection of $\tilde{C}$ into $\MM$, i.e.,
$C\equiv \pi(\tilde{C})$. 
We note that an arbitrary lift of $C$ to $\MP$ is closed.
If $C$ is not contractable to a point, i.e.,
the equivalence class $[C]$ is different from $[e]$,
this offers an example of $\phi_{[C]}=\phi_{[e]}$ with $[C]\ne [e]$.
Accordingly such $[C]$ makes $H$ nontrivial.

\section{Example}
\label{sec:examples}

We examine a slowly modulated periodically driven systems in this
section. Here a modification of the adiabatic theorem is required
for the stationary states that are described by by eigenvectors of 
a Floquet 
operator~\cite{Young-JMP-11-3298,Dranov-JMP-39-1340,Tanaka-JPSJ-80-125002}.
We choose the periodically driven systems
instead of slowly driven Hamiltonian systems because 
the examples in the latter case requires either the divergence or 
crossing of eigenenergies, as is seen in the studies of exotic quantum
holonomy~\cite{Cheon-PLA-248-285,Cheon-PLA-374-144,Cheon-ActaPolytechnica-53-410}.
We refer Ref.~\cite{Tanaka-arXiv-150702827} to apply the present
formulation for adiabatic paths that involves level crossings.

We compare an arbitrary pair $(C_1, C_2)$ of two adiabatic paths 
in a two level system, where we suppose that the absence of
spectral degeneracy in the adiabatic paths. 
After we lay out our result using a parameterization that is
suitable to examine the path topology dependence, of two level systems,
we will show an example of nontrivial pair of paths $(C_1, C_2)$
using a quantum map.

First, we parameterize the adiabatic path using the set of 
eigenprojections $P_1$ and $P_2$
\begin{equation}
  \label{eq:bdef}
  b\equiv\{P_1, P_2\}
  ,
\end{equation}
where the order of the projectors are ignored.
Namely, we will specify a point in the base manifold $\MM$ by $b$.
This amount to the parameterization of adiabatic path by Floquet operator
through
the spectral decomposition
\begin{equation}
  \label{eq:spectralDecomposition}
  {U} = z_1 {P}_1 + z_2 {P}_2,
\end{equation}
where $z_j$ is $j$-th eigenvalue ($j=1,2$), since non-degenerate 
Floquet operator $U$ uniquely specifies $b$. In contrast, there are 
two possible values of the ordered projector $p$
introduced in Eq.~\eqref{eq:p_def}, i.e., $(P_1, P_2)$ and $(P_2, P_1)$.
Note that the definition of $b$ in Eq.~\eqref{eq:bdef} is straightforward
to extend to the systems with an arbitrary number of levels.

Second, we take up a geometric interpretation 
of $b$ and $p$ for the two level system,
utilizing the following parameterization of projection 
operator 
\begin{equation}
  \label{eq:Pparametrize}
  {P}(\bvec{a}) = \frac{1}{2}(1+\bvec{a}\cdot{\paulivec})
  ,
\end{equation}
where $\paulivec$ is the vector consists of Pauli matrices,
and $\bvec{a}$ is a normalized three-dimensional real vector.
The eigenprojectors in Eq.~\eqref{eq:spectralDecomposition} can be 
expressed as ${P}_1=P(\bvec{a})$ and ${P}_2=P(-\bvec{a})$.
Now it is straightforward to see that $p$ and $\bvec{a}$ has
$1:1$ correspondence, which implies that $\MP$ can be identified
with $S^2$. On the other hand, $\pm\bvec{a}$ correspond
to a single point in the $b$-space. Namely, the $b$-space 
can be regarded as $\RPP$, the real projective plane.
Hence the covering map $\pi: S^2\to\RPP$ for the two level system can
be regarded as an identification of the antipodal points in 
the sphere.

Now our argument presented in the previous section is ready to apply.
The fundamental class of the base space
$\pi_1(\RPP)=\set{[e],[\gamma]}$ has two elements, where
$e$ is the closed path that contractable to a point, and
the closed path $\gamma$ is not homotopic to $e$.
On the other hand, our total space $\MP$ is simply connected 
as $\MP=S^2$. Hence $[e]$ and $[\gamma]$, 
the two classes of closed paths, offers two different
monodromy map $\phi_{[e]}$ and $\phi_{[\gamma]}$, 
which correspond to the identity and cyclic permutations of
two eigenprojectors, respectively.

We summarize the analysis of two level systems. When 
the trails of the adiabatic paths $C_1$ and $C_2$ in
$\RPP$ are homotopic, the adiabatic time evolutions of
an eigenprojector along $C_1$ and $C_2$ has no difference.
On the other hand, when $C_1$ are $C_2$ not homotopic, 
the composite closed path $C_2^{-1}\cdot C_1$ is homotopic to
$\gamma$, and the corresponding discrepancy $\phi_{[\gamma]}
=\left(\phi_{[C_2]}\right)^{-1}\circ\phi_{[C_1]}
$ (Eq.~\eqref{eq:compositePhi})
is expressed by the cyclic permutation of two items.

\begin{figure}
  \centering
  \includegraphics[width=4.8cm]{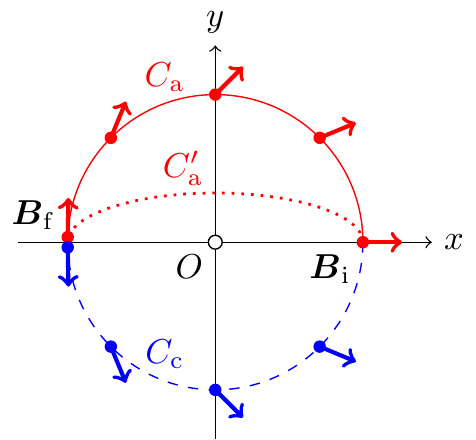}
  \caption{%
    Adiabatic time evolution of $\bvec{a}$ (thick arrow),
    which is equivalent to the ordered set of eigenprojectors
    $p$ for the periodically driven spin-$\frac{1}{2}$
    (Eq.~\eqref{eq:defUspin}).
    Since $\bvec{a}$ is transported from $\bvec{e}_x$
    to $\bvec{e}_y$ along the adiabatic path $C_{\rm a}$ (thick curve), 
    the corresponding
    adiabatic evolution of eigenprojector is from 
    $P(\bvec{e}_x)$ to $P(\bvec{e}_y)$.
    In other words, the adiabatic mapping of $p$ is
    $\phi_{[C_{\rm a}]}\left(p(\bvec{e}_x)\right)
    = p(\bvec{e}_y)$,
    where $p(\bvec{a}) \equiv \left(P(\bvec{a}), P(-\bvec{a})\right)$.
    Other adiabatic path $C_{\rm a}'$ (dotted curve), which is homotopic to 
    $C_{\rm a}$, provides the same adiabatic mapping, 
    i.e., $\phi_{[C_{\rm a}]}= \phi_{[C_{\rm a}']}$.
    On the other hand, the adiabatic path $C_{\rm c}$ (dashed curve)
    is not homotopic to $C_{\rm a}$ due to an obstacle
    (a disclination~\cite{Mermin-RMP-51-591,Nakahara-GTP-1990}) at the origin.
    The corresponding adiabatic evolution is described
    as $\phi_{[C_{\rm c}]}\left(p(\bvec{e}_x)\right)
    = p(-\bvec{e}_y)$.
    We may compare $C_{\rm a}$ and $C_{\rm c}$ by a closed
    path $C=C_{\rm c}^{-1}\cdot{}C_{\rm a}$. 
    The discrepancy between $\phi_{[C_{\rm a}]}$ and 
    $\phi_{[C_{\rm c}]}$ is given by
    $\phi_{[C]}$, which corresponds to the cyclic permutation of
    the two items in $p$.
  }
  \label{fig:example}
\end{figure}

We exemplify the above argument using a slowly modulated driven 
spin-$\frac{1}{2}$, where we set $\hbar=1$.
In the absence of the modulation, our example is described 
by the following periodically driven Hamiltonian:
\begin{equation}
  \label{eq:defHdriven}
  H(t) = 
  \frac{1}{2}\bvec{B}\cdot\paulivec
  + \lambda \frac{1-\sigma_z}{2}\sum_{m=-\infty}^{\infty}
  \delta(t-m)
  ,
\end{equation}
where 
$\bvec{B} = B (\cos\phi\bvec{e}_x+\sin\phi\bvec{e}_y)$ is 
the static magnetic field confined in $xy$-plane
($B$ and $\phi$ are the cylindrical variables),
and $\lambda$ is the strength of the periodic term.
The corresponding Floquet operator is, for example
\begin{equation}
  \label{eq:defUspin}
  U
  = e^{-i\lambda(1-\sigma_z)/2}
  e^{-i\bvec{B}\cdot\paulivec/2}
  ,
\end{equation}
which is a quantum map under a rank-$1$ 
perturbation~\cite{Combesqure-JSP-59-679,Milek-PRA-42-3213}.
In the following, we examine $U$ under the adiabatic changes of
$\bvec{B}$ in $(B_x, B_y)$-plane except the origin $\bvec{B}=0$.
Also, we set $\lambda=\phi$ along the adiabatic path.
Hence $U$ is single-valued in $(B_x,B_y)$-plane,
since 
$U$ periodically depends on $\lambda$ with the period 
$2\pi$.
The eigenvalues of $U$ are, as shown in Ref.~\cite{Tanaka-PLA-379-1693},
$z_{\pm} = \exp\left\{-i(\phi \pm \Delta)/2\right\}$,
where
\begin{align}
  \Delta = 2\arccos\left(\cos\frac{\phi}{2}\cos\frac{B}{2}\right)
  .
\end{align}
The corresponding eigenprojectors are $P(\pm\bvec{a})$, 
where
\begin{align}
  \bvec{a}
  &
  = 
  \frac{1}{\sin(\Delta/2)}
  \left[
    \sin\frac{B}{2}
    \left(\cos\frac{\phi}{2}\bvec{e}_{\rho}
      - \sin\frac{\phi}{2}\bvec{e}_{\phi}
    \right)
    \right.\nonumber \\  &\left.\qquad\qquad\qquad
    - \sin\frac{\phi}{2}\cos\frac{B}{2}\bvec{e}_z\right]
  ,
\end{align}
$\bvec{e}_{\rho}=\cos\phi\bvec{e}_x+\sin\phi\bvec{e}_y$
and $\bvec{e}_{\phi}=-\sin\phi\bvec{e}_x+\cos\phi\bvec{e}_y$.
Note that $\Delta$ and $\bvec{a}$ are not single-valued in
$(B_x,B_y)$-plane although $U$ is single-valued. 
We depict adiabatic paths, whose initial and 
final points in $(B_x, B_y)$-plane are $\bvec{B}_{\rm i} \equiv (\pi, 0)$ 
and $\bvec{B}_{\rm f}\equiv (-\pi,0)$, respectively,
and corresponding adiabatic 
time evolution of eigenspaces in Fig.~\ref{fig:example}. 

\section{Conclusion}
\label{sec:summary}

We have shown the path topology dependence of adiabatic time evolution in closed quantum systems through a topological argument, which is based on the recent study on the exotic quantum holonomy~\cite{Tanaka-PLA-379-1693}. We finally note that examples of systems exhibit non-trivial adiabatic path topology dependence, according to the studies of exotic quantum holonomy in, for example, quantum graphs with generalized connection conditions\cite{Cheon-AP-294-1,Tsutsui-JMP-42-5687,Ohya-AP-331-299,Ohya-AP-351-900},
many-qubit systems~\cite{Tanaka-EPL-96-10005}, adiabatic quantum computation~\cite{Tanaka-PRA-81-022320},
and the Lieb-Liniger model~\cite{Yonezawa-PRA-87-062113}.

\section*{Acknowledgements}
This research was supported by the Japan Ministry of Education, Culture, Sports, Science and Technology under the Grant number 15K05216.



%

\end{document}